 \documentclass[final,1p,times]{elsarticle}

\usepackage{amsmath,amssymb}
\usepackage{graphicx}
\usepackage{wrapfig}

\usepackage[cp1251]{inputenc}

\newcommand{\bea}{\begin{eqnarray}}
\newcommand{\eea}{\end{eqnarray}}
\newcommand{\beaa}{\begin{eqnarray*}}
\newcommand{\eeaa}{\end{eqnarray*}}
\newcommand{\be}{\begin{equation}}
\newcommand{\ee}{\end{equation}}
\newcommand{\bear}{\begin{array}}
\newcommand{\eear}{\end{array}}

\newcommand{\eps}{\varepsilon}

\newcommand{\dtwo}[2]{\frac{\partial^2 #1}{\partial #2^2}}
\newcommand{\done}[2]{\frac{\partial #1}{\partial #2}}
\newcommand{\dthree}[2]{\frac{\partial^3 #1}{\partial #2^3}}

\journal{Physica B}

\begin{document}
\begin{frontmatter}



\title{Defect-mediated relaxation and non-linear susceptibilities of  Rochelle salt}


\author{A.P.Moina}

\address{Institute for Condensed Matter Physics, 1 Svientsitskii Street,
79011, Lviv, \\
 e-mail: alla@icmp.lviv.ua, tel/fax: +380 32
2761158}

\begin{abstract}
The deformable pseudospin Mitsui model is modified in order to
take into account interactions of the ordering dipoles of Rochelle
salt with dipoles, associated with switchable crystal defects.
Using the Glauber-type kinetics of the ordering and defect
pseudospins, we calculate the linear, second, and third order
dynamic susceptibilities and piezoelectric coefficients of the
system. The defect-assisted dispersion of the dynamic
characteristics below 1~kHz is described. Behavior of the linear
and non-linear susceptibilities close to $T_{\rm C1,2}$ is also
satisfactorily described by the presented model.

\end{abstract}

\begin{keyword}
Rochelle salt \sep non-linear susceptibility \sep relaxing defects
\sep rigid defects \sep Mitsui model \sep internal bias field


\end{keyword}

\end{frontmatter}


\section{Introduction}
\label{sect1} Rochelle salt is a curious system, where the
ferroelectric phase exists only in a temperature interval between
two second order phase transitions at 255 and 297~K. Its behavior
is usually described within a two-sublattice Ising model with an
asymmetric double-well potential (Mitsui model \cite{int3,83}) or
its deformable versions
\cite{ourrs,ourrs2,monoclinic,our-diagonal} that take into account
the piezoelectric coupling with the shear strain $\eps_4$ and
diagonal extensional strains $\eps_1$, $\eps_2$, $\eps_3$.
Rochelle salt thus serves as a convenient toy model for a
theoretical exploration of various physical effects in
ferroelectrics with the help of a simple mathematical language,
since already the mean field approximation appears to be
satisfactory here.

Dynamic dielectric response of Rochelle salt exhibits several
dispersions. Those are: related to domain walls motion
\cite{shylnikov} or central thermal peak \cite{Araujo} (below
1~kHz), piezoelectric resonance \cite{Leonovici,muller} (between
10~kHz and 10~MHz), microwave relaxation \cite{int4}, and the
submillimeter (100-700~GHz) resonances \cite{Volkov}. Unruh,
M\"{u}ser,  and others  also observed a Debye-like relaxation of
the dynamic permittivity \cite{Unruh,UnruhSailer} and
piezoelectric coefficient $d_{14}$ \cite{MuserSchmitt} of Rochelle
salt below 1~kHz both in the paraelectric and ferroelectric
phases, which could not be related to the domain-wall motion. It
was found to be strongly dependent on the humidity of the
atmosphere in which the sample were stored and, therefore,
attributed to the influence of lattice defects produced by intake
or loss of crystallization water molecules.

Miga et al \cite{Miga} recently measured the second and third
order dielectric susceptibilities of Rochelle salt. The static
values of these characteristics, calculated within the Mitsui
model, albeit qualitatively correct, are in a severe quantitative
disagreement with the experiment near the Curie temperatures (we
discuss this in detail later). The theoretical curves diverge at
$T_{\rm C1,2}$, whereas in experiment the anomalies of the
susceptibilities are lowered down and smeared out. It is generally
known that the behavior of the physical characteristics of
ferroelectrics in the transition regions is strongly affected by
the presence of defects in the crystals. Hence, the
above-mentioned relaxation below 1~kHz and the observed smearing
of the susceptibilities anomalies can be of the same origin and
 attributed to the defect-induced fields and
defect-assisted relaxation in the system.

In the present paper we develop a model that describes both the
low-frequency relaxation and the behavior of the linear and
non-linear susceptibilities in Rochelle salt. The paper is
organized as follows. In Section 2 a short review of the
literature on the notion of the defect-induced intrinsic field in
ferroelectrics is given. In Section~3 the model is formulated, and
its static thermodynamic properties are calculated. In Section~4
we consider dynamics of the system and obtain expressions for the
linear and non-linear dynamic susceptibilities and piezoelectric
coefficients of Rochelle salt. Numerical calculations are
performed in Section~5, and concluding remarks are presented in
Section~6.

\section{Defect-associated fields. Switchable defects}
Ferroelectric crystals may have various  defects. We shall deal
here mostly with the dipole or polarized  defects that cannot
migrate over a crystal, but can be reoriented (switched) by
external electric field or relax thermally. The notion of the
defect-induced internal bias field relies on the assumption that
switchable or relaxing defects give rise to a bias field $E_e$,
which direction always coincide with the direction of polarization
(crystal polarization in the case of a single domain crystal or
with domain polarization in the multi-domain case) and which
magnitude is proportional to the value of polarization
\cite{Unruh}
\begin{equation}
\label{eingepragte} \eps_0E_e=AP.
\end{equation}
The linear correlation between the internal bias field $E_e$ of
switchable defects and spontaneous polarization  has been
experimentally confirmed, for instance, for Rochelle salt
\cite{Unruh}, $\gamma$-irradiated TGS \cite{Hilczer}, and lossy
KH$_2$PO$_4$ \cite{Abe-defects}. Temperature variation of the
parameter $A$ in Rochelle salt and its dependence on the value of
atmosphere humidity, in which the samples were stored for
sufficiently long periods of time, has been explored in
\cite{UnruhSailer}. Arlt et al \cite{Arlt2} found the parameter
$A$ to be inversely proportional to the dielectric constant of the
host ferroelectric material, but their calculations did not take
into account the converse effect: a strong dependence of the
dielectric constant of the ferroelectrics on the electric field at
temperatures close to the Curie point.

Dynamics of the bias field $E_e$ is relaxational, most easily
described  by the equation \cite{Unruh}
\begin{equation}
\label{Ee} -\tau\frac{{\rm d} E_e}{\rm d t}=E_e-E_{e0},\quad
E_{e0}=\frac{A}{\eps_0}P.
\end{equation}
The quantity $E_{e0}$, towards which the field $E_e$ is relaxing,
is proportional to the momentary value of polarization $P$.
Matsubara et al \cite{Matsubara-defects} considered, instead of
$E_e$, motion of defects in a two-well potential. For the
difference between populations of the two wells they obtained an
equation similar to Eq.~(\ref{Ee}), and the analog of $A$ was
found to be  inversely proportional to temperature. Existence of
relaxing internal bias fields directed along the domain
polarization explains, for instance, an anomalous temperature
behavior of the coercive field in lossy KH$_2$PO$_4$
\cite{Abe-defects,Matsubara-defects}, as well as the transient
double hysteresis loops in various defective ferroelectrics.

 The relaxation time $\tau$ has been found \cite{Unruh,UnruhSailer} to have an
Arrhenius behavior $\tau=\tau_0\exp(W/k_{\rm B}T)$. The activation
energy $W$ in Rochelle salt varied between 0.4 and 0.8~eV.

\section{The model}
\label{model}

The system we consider consists of i) ordering dipoles; ii)
switchable defect dipoles, iii) rigid defects, and iv) host
lattice.

The ordering dipoles are those responsible for the phase
transitions and formation of spontaneous polarization in the
crystal. They are described by the deformable Mitsui model
\cite{ourrs,our-diagonal}, which considers motion of pseudospins
$\sigma_{qf}=\pm1$ in two interpenetrating sublattices $f=1,2$
with asymmetric double well potentials and their interactions to
the lattice strains and electric fields; $q$ is the unit cell
index.

The switchable defect dipoles are believed to be trapped on
specific sites within a given unit cell. The switching is a
jump-like process between two potential wells. Thus, the
orientation of a dipole sitting on the site $i$ in the  $q$-th
unit cell can be described by the pseudospin operator
$S_{qi}=\pm1$. In the case of Rochelle salt the switchable defects
are, most likely, the dipoles formed by water vacancies or
interstitials, for samples stored in highly dry or wet atmosphere,
respectively.

Rigid dipole defects that cannot be reoriented (or if their
reorientation is so slow that it can be ignored on the time scales
of the motions of switchable dipoles and of ordering dipoles) are
assumed to create a constant bias field $E_b$, directed along the
axis of spontaneous polarization (100) and proportional to the
concentration of these defects.  The transverse components of this
field are ignored, and it is taken to be temperature independent.
The rigid dipoles can be formed, for instance, by impurity-vacancy
complexes, like those observed in doped Rochelle salt
\cite{be-doped} and having the relaxation times of the order of 10
min at $T_{\rm C2}$ and 10$^3$~min at $T_{\rm C1}$. Another option
is that screw dislocations are the source of the constant bias,
creating around them a shear stress $\sigma_{4b}$. This stress,
just like the longitudinal electric field $E_1$, induces
polarization $P_1$ and shear strain $\eps_4$. Since the action of
$E_1$ and $\sigma_4$ is equivalent, we can describe the influence
of rigid dipoles either via $E_b$ or via $\sigma_{4b}$.

Strictly speaking, if the external bias field conjugate to the
order parameter is applied, the second order phase transitions in
the system are smeared out. Physical characteristics of the
system, such as the dielectric susceptibility or piezoelectric
coefficient associated with the order parameter, then have only
rounded maxima at temperatures close to the Curie temperatures of
a crystal, not placed in a bias field. Nevertheless, we shall call
the temperatures of these maxima the Curie temperatures $T_{\rm
C1}$ and $T_{\rm C2}$, remembering that those are not truly second
order phase transitions.

 The total Hamiltonian of the system will be written
in the following form
\begin{equation}
\label{totalHam} H=H_\sigma+H_S+H_{int}+NU_{seed},
\end{equation}
where $H_\sigma$ is the Hamiltonian of the modified Mitsui model
\cite{ourrs,our-diagonal}.
\begin{equation}
H_\sigma=-\frac12 \sum\limits_{qq'}\sum\limits_{ff'=1}^2
R_{qq'}^{ff'} \frac{\sigma_{qf}}{2}\frac{\sigma_{q'f'}}{2} -\Delta
\sum\limits_q \left(\frac{\sigma_{q_1}}{2} -\frac{\sigma_{q_2}}{2}
\right)- [\mu_1(E_1+E_b)-2\psi_4\varepsilon_4]
\sum\limits_q\sum\limits_{f=1}^2 \frac{\sigma_{qf}}{2}.
\end{equation}
Here the parameter $\Delta$ describes the asymmetry of the double
well potential; $\mu_1$ is the effective dipole moment of the
ordering pseudospins. The model parameter $\psi_4$ describes the
internal field created by the piezoelectric coupling with the
shear strain $\varepsilon_4$; $E_1$ is an external longitudinal
electric field. $R_{qq'}^{11} = R_{qq'}^{22} = J_{qq'}$ and
$R_{qq'}^{12} = R_{qq'}^{21} = K_{qq'}$ are the potentials of
interaction between the ordering pseudospins, belonging to the
same and to different sublattices, respectively.

The second and third terms in Eq.~(\ref{totalHam}) describe
interactions of the defect dipoles with the external and constant
bias electric fields, their coupling to the shear strain $\eps_4$
\[
H_S=-\sum_{qi}\left[
m_1(E_1+E_b)-2\Psi_4\eps_4\right]\frac{S_{qi}X_{qi}}{2},
\]
and to the ordering pseudospins $\sigma_{qf}$
\[
H_{int}=-\sum\limits_{qq'}\sum_{f=1}^2\sum_i \lambda^{fi}_{qq'}
\frac{\sigma_{qf}}{2}\frac{S_{q'i}X_{q'i}}{2}.
\]
Here summation over $i$ is carried out over the sites that can be
occupied by defects in a given cell; $m_1$ is the dipole moment of
a defect dipole; $X_{qi}=1$ if the defect dipole site is occupied,
and $X_{qi}=0$ otherwise. $\sum_i\langle X_{qi}\rangle=c$ is the
concentration of the defect dipoles: the average number of defects
per unit cell (two formula units of Rochelle salt). It is assumed
to be small, so the interactions between the defect dipoles, which
would be proportional to $c^2$, are not considered.

Finally, the phenomenological part of the Hamiltonian  $NU_{seed}$
is a ``seed'' energy of the host lattice of heavy ions which forms
the asymmetric potentials for the ordering pseudospins
\begin{equation}
{U_{seed}}=\frac{v}{2} c_{44}^{E0}\varepsilon_4^2
-{v}e_{14}^0\varepsilon_4  (E_1+E_b) - \frac{v \eps_0}{2}
\chi_{11}^{\varepsilon0} (E_1+E_b)^2 + \frac{v}{2}
\sum_{i,j=1}^3c_{ij}^{E0}\varepsilon_i\eps_j
-v\sum_{ij=1}^3c_{ij}^{E0}\alpha_i^0(T-T_i^0)\eps_j.
\end{equation}
Here  $N$ is the number of the unit cells; $\eps_0$ is the vacuum
permittivity; $v$ is the unit cell volume of the model;
$c_{44}^{E0}$, $c_{ij}^{E0}$, $e_{14}^0$, $\alpha_i^0$ are the
``seed'' constants describing the phenomenological contributions
of the crystal lattice into the corresponding observed quantities.

Using the  mean field approximation, we obtain the following
expression for the thermodynamic potential of the system (per one
unit cell)
 \begin{eqnarray}
 \label{pot}
&& g_{2E}(\sigma_i,T) =-v\sum_{i=1}^4\sigma_i\eps_i+U_{ seed}  -
\frac{2\ln 2}{\beta}+ \frac{J +K}4\xi^2 + \frac{J
-K}4\sigma^2+c\lambda S\xi
 \\
 && -  \frac1\beta\ln \cosh
\frac{\gamma +\beta\lambda c S+ \delta}2
 \cosh \frac{\gamma+\beta\lambda c
S - \delta}2-\frac{c}{\beta}\ln
\cosh\beta\frac{2\lambda\xi-2\Psi_4\eps_4+m_1(E_1+E_b)}{2},
\nonumber
\end{eqnarray}
where $\beta=1/k_{B}T$, $k_B$ is the Boltzmann constant,
$\sigma_i$ are the components of the elastic stress tensor, and
\begin{equation}
\label{gammadelta} \gamma = \beta\left[ \frac{J + K}{2}\xi -
2\psi_4\varepsilon_4 + \mu_1(E_1+E_b) \right], \quad \delta =
\beta \left( \frac{J - K}{2}\sigma + \Delta\right).
\end{equation}

Here $J$, $K$, $\lambda$ are the Fourier-transforms (at ${\bf
k}=0$) of the constants of interaction between the ordering and
defect pseudospins. $J$ and $K$, along with the asymmetry
parameter $\Delta$, are taken to be linear functions of the
diagonal strains \cite{monoclinic,our-diagonal}
 \begin{equation}\label{2.2}
    J \pm K= J_0 \pm K_0+2 \sum\limits_{i = 1}^3 {\psi_{i}^\pm \varepsilon _i
    },\quad \Delta = \Delta _0 + \sum\limits_{i = 1}^3 {\psi_{3i}
\varepsilon _i }. \end{equation}
For $J$ and $K$ such an expansion
is equivalent to taking into account the electrostrictive coupling
with the diagonal strains.

The system behavior is described in terms of the  mean pseudospin
values
\begin{equation}
\xi = \frac{\langle\sigma_{q1}\rangle+
\langle\sigma_{q2}\rangle}2, \quad \sigma =
\frac{\langle\sigma_{q1}\rangle -\langle\sigma_{q2}\rangle}2,\quad
S=\langle S_{qi}\rangle \label{gammsigma}
\end{equation}
$\xi$ is the parameter of ferroelectric ordering in the system.
They  are determined from the saddle point of the thermodynamic
potential (\ref{pot}): a minimum of $g_{2E}$ with respect to $\xi$
and $S$ and a maximum with respect to $\sigma$ are realized at
equilibrium. The corresponding equations can be written as
\begin{eqnarray}
&& \xi = \frac12[\tanh \frac{\gamma+\beta\lambda c
S+\delta}2+\tanh\frac{\gamma+\beta\lambda c S-\delta}2],
\nonumber\\
 &&\sigma =
 \frac12[\tanh
\frac{\gamma+\beta\lambda c
S+\delta}2-\tanh\frac{\gamma+\beta\lambda c S-\delta}2],\nonumber\\
&&\label{ord-par} S =
 \tanh
\beta\frac{2\lambda\xi-2\Psi_4\eps_4+m_1(E_1+E_b)}{2}.
\end{eqnarray}
Note that in the thermodynamic potential (\ref{pot}) and in
Eq.~(\ref{gammsigma}) $\xi$, $\sigma$, $S$, and $c$ are taken to
be independent of the unit cell index $q$, i.e. the spatial
fluctuations of the defect concentration and of the pseudospin
mean values are ignored.

The stress-strain relations and polarization  are derived from the
thermodynamic potential
\begin{eqnarray}
\label{2.4} && \sigma_i = \frac{1}{\bar v} \left( \frac{\partial
g_{2E}}{\partial \varepsilon_i} \right)_{E_1,\sigma_i}=
 \sum\limits_{j = 1}^3 c_{ij}^{E0} [\varepsilon _j
-\alpha_j^0(T-T_j^0)]
- \frac{1}{2v}\psi_i^+  \xi ^2 - \frac{1}{2v}\psi_i^- \sigma ^2
- \frac{1}{v}\psi_{3i} \sigma, \quad ({i } = 1-3)\nonumber \\
\label{2.4a} && \sigma_4 = c_{44}^{E0} \varepsilon _4 - e_{14}^0
(E_1+E_b) +
2\frac{\psi _4 }{v}\xi+c\frac{\Psi _4 }{v}S, \nonumber\\
 \label{2.4b} && P_1 = -\frac{1}{\bar v} \left( \frac{\partial g_{2E}}{\partial E_1}
\right)=e_{14}^0 \varepsilon _4 + \chi _{11}^{\varepsilon
0}(E_1+E_b) + \frac{\mu _1 }{v}\xi + c\frac{m _1 }{2v}S.
 \end{eqnarray}
Linearizing the last of Eq.~(\ref{ord-par}) and substituting the
result into the two first equations, one can see that coupling to
the defect dipoles is equivalent to appearance of an additional
field $E_e$, acting on the ordering dipoles
\begin{equation}\label{eingepragte-feld}
E_e=\frac{\beta c\lambda^2 \xi}{\mu_1}.
\end{equation}
It is  inversely proportional to temperature and proportional to
the order parameter $\xi$ and, if we neglect all the contributions
into polarization Eq.~(\ref{2.4b}) other than due to $\xi$, also
to the polarization $P_1$. In this case we can relate parameters
of our model to the constant $A$ of Eq.~(\ref{eingepragte}),
introduced by Unruh et al \cite{Unruh}, as
\begin{equation}
\label{a} A=\frac{v\eps_0 \beta c\lambda^2}{\mu_1^2}.
\end{equation}


%
%

\section{Dynamic linear and non-linear susceptibilities of Rochelle salt}
We  consider a dielectric and piezoelectric response of a thin
rectangular $l_y\times l_z$ plate of a Rochelle salt crystal cut
in the (100) plane (0$^\circ$ X-cut, the sample edges parallel to
[010] and [001]), induced by a time-dependent harmonic electric
field $E_{1t}\exp(i\omega t)$. This field gives rise to the shear
strain $\eps_4$ at all temperatures, as well as to the diagonal
strains $\eps_1$, $\eps_2$, $\eps_3$ in the ferroelectric phase.
Influence of the in-plane extensional vibrational modes associated
with $\eps_2$ and $\eps_3$ on the dynamic permittivity of Rochelle
salt X-cuts has been explored in detail in \cite{PhysB}. In
particular, it was shown that the extensional modes are excited
only in the ferroelectric phase (as follows from the system
symmetry), and that the lowest piezoelectric resonance frequency
is always associated with the shear mode. For the sake of
simplicity, in the present consideration the dynamics of the
diagonal strains will be ignored.

  Dynamics of the
strain $\eps_4$ will be described, using classical (Newtonian)
equations of motion \cite{Masonbook} of an elementary volume
\begin{equation}
\label{11} \rho\dtwo{\eta_i}{t}=\sum_k \done{\sigma_{ik}}{x_k},
\end{equation}
where $\rho=1767$~kg/m$^3$ is the crystal density; $\eta_i$ are
displacements of an elementary volume along the axis $x_i$;
$\sigma_{ik}$ are components of the stress tensor. From here one
easily derives that
\begin{equation}
\label{23}
\rho\dtwo{\eps_4}t=\dtwo{\sigma_{4}}{y}+\dtwo{\sigma_{4}}{z}.
\end{equation}

Dynamics of the ordering and defect pseudospins will be described
within the Glauber approach \cite{int9}. The  kinetic equations
for the time-dependent variables $\xi$ and $\sigma$, associated
with the ordering pseudospins, read \cite{ourrs}
\begin{eqnarray}
-\alpha\frac{d}{dt}\xi=\xi-\frac{1}{2}[\tanh\frac12(\gamma+\beta\lambda
cS+\delta)+ \tanh\frac12(\gamma+\beta\lambda cS-\delta)],
\nonumber\\
\label{3.3}
-\alpha\frac{d}{dt}\sigma=\sigma-\frac{1}{2}[\tanh\frac12(\gamma+\beta\lambda
cS+\delta)- \tanh\frac12(\gamma+\beta\lambda cS-\delta)].
\end{eqnarray}
Here $\alpha$  is the parameter setting the time scale of this
dynamics; its value is usually found by fitting theoretical curves
of the permittivity in the  microwave frequency range to
experiment \cite{ourrs,ourrs2,our-comp}.

For dynamics of the defect pseudospins a similar equation is
obtained
\begin{equation}
\label{24}
-\tau\frac{d}{dt}S=S-\tanh\beta\frac{2\lambda\xi-2\Psi_4\eps_4+m_1(E_1+E_b)}{2},
\end{equation}
however, with a different time scale parameter $\tau$.

These equations, in fact, describe three different dynamic
phenomena: the intrinsic dynamics of the pseudospin subsystem,
expected to occur at microwave frequencies, the strain dynamics,
yielding the piezoelectric resonances, and the defect-mediated
relaxation, expected to occur below 10~kHz. We shall show that all
three processes take place in well separated frequency ranges, in
particular, that the piezoelectric resonances do not overlap with
the defect-mediated relaxation.

We present the dynamic variables $\xi$, $\sigma$, $\eps_4$, $S$,
and their linear functions $\gamma$ and $\delta$,
Eq.~(\ref{gammadelta}) as sums of the equilibrium values and of
the fluctuational deviations, while the deviations are taken to be
in the form of harmonic waves, e.g.
 \[
\xi=\xi^{(0)}+\sum_{n}\xi^{(n)}(y,z)\exp(in\omega t), \quad
\xi^{(n)}(y,z)\sim E_{1t}^n,
\]
etc. Fluctuations of the diagonal strains $\eps_1$,  $\eps_2$,
$\eps_3$ are neglected.

Equations (\ref{2.4}), (\ref{23})--(\ref{24}) are expanded in
these deviations up to the cubic in $E_{1t}$ terms. Then in these
equations the terms proportional to  the same power of $E_{1t}$
are collected. For the equilibrium quantities we obtain equations
(\ref{ord-par}) and (\ref{2.4}) with $\xi$, $\sigma$, $\eps_i$,
$S$ replaced with their equilibrium values $\xi^{(0)}$, $\sigma$,
$\eps_i$, $S$.

As the constitutive equations  are linear, their fluctuation parts
of the order of $E_{1t}^n$ for each $n\geq1$ can be written as
\begin{eqnarray}
 && \sigma_4^{(n)}(y,z) = c_{44}^{E0} \varepsilon
_{4}^{(n)}(y,z) - e_{14}^0 E_{1t}\delta_{n,1} +
2\frac{\psi _4 }{v}\xi^{(n)}(y,z) +c\frac{\Psi _4 }{v}S^{(n)}(y,z),\nonumber \\
 \label{2.4bf} && P_{1}^{(n)}(y,z) = e_{14}^0 \varepsilon
_{4}^{(n)}(y,z) + \chi _{11}^{\varepsilon 0} E_{1t}\delta_{n,1} +
\frac{\mu _1
}{v}\xi^{(n)}(y,z)+c\frac{m_1}{2v}S^{(n)}(y,z)\label{2.4af} .
 \end{eqnarray}
$\delta_{n,1}$ is the Kronecker symbol.

Equations for the strain (\ref{23})  are linear too, yielding
\begin{equation}\label{strain-all}
    -\rho(n\omega)^2\eps_{4}^{(n)}=
c_{44}^{E0}\left(\dtwo{\eps_{4}^{(n)}}{z}+\dtwo{\eps_{4}^{(n)}}{y}\right)+
\frac{2\psi_4}{
v}\left(\dtwo{\xi^{(n)}}{z}+\dtwo{\xi^{(n)}}{y}\right)
+c\frac{\Psi_4}{
v}\left(\dtwo{S^{(n)}}{z}+\dtwo{S^{(n)}}{y}\right).
\end{equation}
We shall also linearize Eq.~(\ref{24}), thus
\begin{equation}
\label{Ee-all}
S^{(n)}=\frac{\beta}{2}\frac{2\lambda\xi^{(n)}-2\Psi_4\eps_4^{(n)}+m_1E_{1t}\delta_{n,1}}{1+in\omega\tau}.
 \end{equation}
Kinetic equations (\ref{3.3}) are non-linear and remain so, hence
their form is different for different $n$.

\subsection{Linear characteristics}
Linear in $E_{1t}$ part of Eqs.~(\ref{3.3}) reads
\begin{eqnarray}
  &&-\xi^{(1)}(1+i\alpha\omega) + c_2^+\left(\gamma^{(1)}+\beta\lambda c S^{(1)}+\delta^{(1)}\right)+
  c_2^-\left(\gamma^{(1)}+\beta\lambda c S^{(1)}-\delta^{(1)}\right)=0,\nonumber \\
&&-\sigma^{(1)}(1+i\alpha\omega) +
c_2^+\left(\gamma^{(1)}+\beta\lambda c
S^{(1)}+\delta^{(1)}\right)-
  c_2^-\left(\gamma^{(1)}+\beta\lambda c S^{(1)}-\delta^{(1)}\right)=0,\label{3.3first}
\end{eqnarray}
where
\begin{equation}\label{c2}
    c_2^\pm=\frac14[1-\tanh^2\frac{\gamma^{(0)}+\beta\lambda c S^{(0)}\pm\delta^{(0)}}{2}].
\end{equation}
We shall be mostly interested here in the system behavior in the
frequency range below 10~kHz. At these frequencies and with the
value of $\alpha\sim 10^{-13}$c$^{-1}$
\cite{ourrs,ourrs2,our-comp} chosen to describe the microwave
relaxation in Rochelle salt, the terms proportional to
$\alpha\omega$ are negligibly small and shall be omitted. The
intrinsic dynamics of the ordering pseudospin subsystem becomes
irrelevant. The value of $\tau$, on the other hand, will be chosen
to describe the  possible dispersion of the permittivity below
10~kHz,  caused by dynamics of the defect dipoles
(Eq.~(\ref{24})).

From Eqs. (\ref{2.4af}), (\ref{Ee-all})  and (\ref{3.3first}) at
$n=1$ we find
\begin{eqnarray}
&&\label{xi-sigma-eps}\xi^{(1)}(y,z)=
\frac{\beta[\mu_1+\Delta_\mu(\omega)]}{2}F_1(\omega)E_{1t}-\beta[\psi_4+\Delta_\psi(\omega)]
F_1(\omega)\eps_{4}^{(1)}(y,z)
\nonumber,\\
&&  \sigma^{(1)}(y,z)=
\frac{\beta[\mu_1+\Delta_\mu(\omega)]}2F_1^\sigma(\omega)E_{1t}-\beta[\psi_4+\Delta_\psi(\omega)]
F_1^\sigma(\omega)\eps_{4}^{(1)}(y,z),
\end{eqnarray}
 Here
\begin{eqnarray} &&F_1(n\omega)=\frac{\varphi_3}{\varphi _2-\Lambda_e(n\omega)\varphi_3},\nonumber\\
&&
\Lambda_e(n\omega)=\frac{1}2\frac{\beta^2 c\lambda^2}{1+in\omega\tau},\nonumber\\
&&
\Delta_\psi(n\omega)=\frac{1}2\frac{\beta c\lambda\Psi_4}{1+in\omega\tau},\nonumber\\
&&
 \label{renormalization}
\Delta_\mu(n\omega)=\frac{1}2\frac{\beta c\lambda
m_1}{1+in\omega\tau}.
 \end{eqnarray}
Substituting  Eq.~(\ref{xi-sigma-eps}) into Eq.(\ref{strain-all}),
we obtain an equation for the strain $\eps_{4}^{(1)}(y,z)$
\begin{eqnarray}
&& \label{shortsystem} -\rho\omega^2\eps_{4}^{(1)}=
\tilde{c}_{44}^{E}(\omega)\left[\dtwo{\eps_{4}^{(1)}}{y}+\dtwo{\eps_{4}^{(1)}}{z}\right].
\end{eqnarray}
The boundary condition follows from the assumption that the
crystal is traction free at its edges (at $y=0$, $y=l_y$, $z=0$,
$z=l_z$, to be denoted as $\Sigma$): $\sigma_4|_\Sigma=0$. Using
the constitutive equations we get
\begin{equation}\label{boundary2}
\eps_{4}^{(1)}|_\Sigma=d_{14}^{(1)}(\omega)E_{1t},
\end{equation}
with linear dynamic piezoelectric coefficients and elastic
constant given by
\begin{eqnarray}&& \label{d14} d_{14}^{(1)}(\omega) =
\frac{e_{14}(\omega)}{ c_{44}^E(\omega)},\\
&& e_{14}(\omega)=e_{14}^{0}+\Delta_e(\omega)-
\frac{\beta[\mu_1+\Delta_\mu(\omega)][\psi_4+\Delta_\psi(\omega)]}{v}F_1(\omega),\nonumber\\
&&
{c}_{44}^{E}(\omega)=c_{44}^{E0}+\Delta_C(\omega)-\frac{2\beta[\psi_4+\Delta_\psi(\omega)]^2}{v}F_1(\omega).\nonumber
 \end{eqnarray}
 Here \[\Delta_e(n\omega)=\frac{1}{2v}\frac{m_1c \beta\Psi_4}{1+in\omega\tau}
,\quad \Delta_C(n\omega)=-\frac{1}{v}\frac{c\beta
\Psi_4^2}{1+in\omega\tau}.
\]
  A solution of Eq. (\ref{shortsystem}) with the boundary condition (\ref{boundary2})
  can be written as
\begin{equation}
\eps_{4}^{(1)}=d_{14}^{(1)}(\omega)E_{1t}\left[1+\sum_{kl}\frac{16}{\pi^2(2k+1)(2l+1)}
\frac{\omega^2}{\omega_{kl}^2-\omega^2}\sin\frac{\pi (2k+1)
y}{l_y}\sin\frac{\pi (2l+1) z}{l_z}\right], \label{series}
\end{equation}
where
 $\omega_{kl}$ are given by equation
\begin{equation}
\label{res4} \omega_{kl}=\sqrt{\frac{
c_{44}^E(\omega_{kl})\pi^2}{\rho}\left[\frac{(2k+1)^2}{l_y^2}+\frac{(2l+1)^2}{l_z^2}\right]}.
\end{equation}

The observable linear dynamic dielectric susceptibility is
expressed via the derivative from the polarization averaged over
the sample volume
\begin{eqnarray}
\label{permittivity}
&&\chi_{11}^{(1)}(\omega)=\frac1{l_yl_z}\frac{1}{\eps_0}\frac{\partial{}}{\partial
E_{1t}} \int_0^{l_y} dy\int_0^{l_z} dz
P^{(1)}_{1}(y,z)\nonumber\\&&\quad
=\chi_{11}^{\eps0}+\frac{\beta\mu_1^2}{2v\eps_0}F_1(\omega)+
e_{14}(\omega)d_{14}^{(1)}(\omega)R_4(\omega),
\end{eqnarray}
with
\[
R_4(\omega)=1+\sum_{k,l}\frac{64}{[\pi^2(2k+1)(2l+1)]^2}
\frac{\omega^2}{\omega_{kl}^2-\omega^2}.
\]
It has a resonance dispersion with peaks at frequencies where
${\rm Re}[R_4(\omega)]\to\infty$.

\begin{figure}[htb]
\centerline{\includegraphics[width=0.35\textwidth]{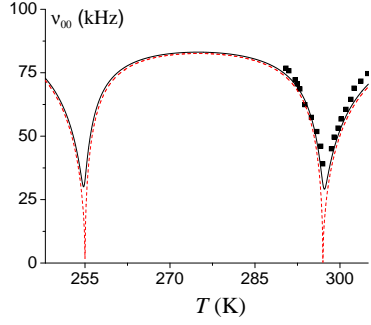}}
\caption{Temperature dependences of the lowest resonance frequency
of the Rochelle salt X-cut with $l_y=1.60$~cm, $l_z=2.45$~cm .
Solid line: the present theory with the defect-mediated
relaxation, $A_0=1.1\cdot 10^{-4}$, $E_b=4.5 \cdot 10^3$~V/m,
$\tau_0=5.95\cdot 10^{-16}$~s, $W=0.5$~eV; dashed line: the
earlier theory \cite{PhysB}. Symbols: experimental points of
\cite{Leonovici}. The choice of the values of $A_0$ and other
model parameters is discussed in Section~\ref{NA}.} \label{lowres}
\end{figure}

Figure~\ref{lowres} shows the lowest resonant frequency
$\nu_{00}=\omega_{00}/2\pi$ of a rectangular Rochelle salt X-cut,
calculated within the model without the defect-mediated relaxation
\cite{PhysB} and within the present model. Both theories yield
identical results for all temperatures except for the very narrow
regions around the Curie temperatures. At the transition points
the resonant frequency goes to zero in an ideal crystal, whereas
in a crystal with defects they do not drop below 10~kHz even for a
relatively large sample with $l_y=1.60$~cm, $l_z=2.45$~cm.

Hence, in the frequency range of interest ($\nu \lesssim 1$~kHz)
we have $\omega_{kl}^2\gg \omega^2$,  and $R_4(\omega)\approx 1$.
Therefore, we can ignore the spatial variation of the dynamical
variables $\xi$, $\sigma$, $\eps_4$. In this case
$\eps_4^{(1)}\approx d_{14}^{(1)}(\omega)E_{1t}$; the linear
dynamic piezoelectric coefficient $d_{14}^{(1)}(\omega)$ is given
by
\begin{equation}
\label{dyn_d14}
d_{14}^{(1)}(\omega)=\frac{e_{14}^0+\Delta_e(\omega)}{c_{44}^{E0}+\Delta_C(\omega)}-\frac{\beta\mu'_1(\omega)
}{v[c_{44}^{E0}+\Delta_C(\omega)]}[\psi_4+\Delta_\psi(\omega)]N(\omega)\varphi_3,
\end{equation}
whereas the linear susceptibility reads
\begin{equation}
\label{dyn_susc}
\chi_{11}^{(1)}(\omega)=\chi_{11}^{\sigma0}(\omega)+\frac{\beta[\mu'_1(\omega)]^2}{2v\eps_0}N(\omega){\varphi_3},
\end{equation}
where
\begin{eqnarray}\label{N}
 &&
 N(\omega)=\frac{1}{\varphi_2-\Lambda_e(\omega)\varphi_3-\Lambda(\omega)\varphi_3},\nonumber\\
&&
\mu'_1(\omega)=\mu_1+\Delta_\mu(\omega)-2\frac{e_{14}^0+\Delta_e(\omega)}{c_{44}^{E0}+\Delta_C(\omega)}
 [\psi_4+\Delta_\psi(\omega)],\nonumber\\
&& \label{renormalization2}
\Lambda(\omega)=\frac{2\beta[\psi_4+\Delta_\psi(\omega)]^2}{v[c_{44}^{E0}+\Delta_C(\omega)]},\nonumber\\
&&\chi_{11}^{\sigma0}(\omega)=\chi_{11}^{\eps0}+c\frac{\beta
m_1^2}{4v\eps_0}+\frac{[e_{14}^0+\Delta_e(\omega)]^2}{c_{44}^{E0}+\Delta_c(\omega)}.
\end{eqnarray}

As one can see, taking into account the influence of switchable
defect dipoles led to a frequency-dependent renormalization of
almost all constants of the Mitsui model (see Eqs.
(\ref{renormalization}), (\ref{renormalization2})). However, the
role of this renormalization in the system dynamics is minor. The
defect-mediated relaxational dispersion of the linear
susceptibility $\chi_{11}^{(1)}(\omega)$ and piezoelectric
coefficient $d_{14}^{(1)}(\omega)$ is mostly caused by the term
$\Lambda_e(\omega)\varphi_3$ in the denominator of $N(\omega)$.
The susceptibility dispersion width is roughly given by the
expression
\begin{equation}
\label{dispersionwidth} \Delta
\chi_{11}^{(1)}\approx\frac{v\eps_0}{\mu_1^2}\beta
c\lambda^2[\chi_{11}^{(1)}(\infty)]^2=A[\chi_{11}^{(1)}(\infty)]^2,
\end{equation}
where $A$ is given by Eq.~(\ref{a}), and $\chi_{11}^{(1)}(\infty)$
is the susceptibility at frequencies above the defect-mediated
dispersion but below the piezoelectric resonances. The dispersion
width strongly increases as temperature approaches the transition
points. No dispersion is present if $A=0$, i.e. without
interactions between ordering and switchable defect dipoles.

\subsection{Non-linear susceptibilities}
Quadratic in $E_{1t}$ part of the kinetic equations (\ref{3.3})
reads
\begin{eqnarray}
  &&-\xi^{(2)} + c_2^+(\gamma^{(2)}+\beta\lambda c S^{(2)}+\delta^{(2)})+
  c_2^-(\gamma^{(2)}+\beta\lambda c S^{(2)}-\delta^{(2)})\nonumber\\
&&  \qquad +
  c_3^+(\gamma^{(1)}+\beta\lambda c S^{(1)}+\delta^{(1)})^2+
  c_3^-(\gamma^{(1)}+\beta\lambda c S^{(1)}-\delta^{(1)})^2=0,\nonumber \\
&&-\sigma^{(2)} + c_2^+(\gamma^{(2)}+\beta\lambda c
S^{(2)}+\delta^{(2)})-
  c_2^-(\gamma^{(2)}+\beta\lambda c S^{(2)}-\delta^{(2)})\nonumber\\
&&  \qquad +
  c_3^+(\gamma^{(1)}+\beta\lambda c S^{(1)}+\delta^{(1)})^2-
  c_3^-(\gamma^{(1)}+\beta\lambda c S^{(1)}-\delta^{(1)})^2=0,\label{3.3second}
\end{eqnarray}
where
\[
c_3^\pm=\frac{1}{8}\left[\tanh\frac{\gamma^{(0)}+\beta\lambda c
S^{(0)} \pm\delta^{(0)}}{2}-\tanh^3\frac{\gamma^{(0)}+\beta\lambda
c S^{(0)}\pm\delta^{(0)}}{2}\right].\] The spatial variation of
the strain $\eps_4^{(2)}$ is neglected, and the strain is found
from the constitutive equations (\ref{2.4af}) at
$\sigma_4^{(n)}=0$, instead of Eq.~(\ref{strain-all}).

Using the obtained in the previous subsection $\xi^{(1)}$,
$\sigma^{(1)}$, $\eps_4^{(1)}$, and $S^{(1)}$ to find
$\gamma^{(1)}$ and $\delta^{(1)}$ (see Appendix), we solve the
system of equations (\ref{2.4af})-(\ref{Ee-all}) and
(\ref{3.3second}) with respect to the second order quantities
$\xi^{(2)}$, $\eps_4^{(2)}$, etc, and from whence obtain the
second order dynamic dielectric susceptibility
\begin{eqnarray}
&&\chi_{111}^{(2)}(\omega)=\frac1{2\eps_0}\dtwo{P_1^{(2)}}{E_{1t}}=
-\frac{\beta^2[\mu_1'(\omega)]^2\mu_1'(2\omega)}{4v\eps_0}N^2(\omega)N(2\omega)K^{(2)}
\end{eqnarray}
and piezoelectric coefficient
\begin{eqnarray}
&&d_{114}^{(2)}(\omega)=\frac12\dtwo{\eps_4^{(2)}}{E_{1t}}=
\frac{\beta^2[\mu_1'(\omega)]^2}{2v}\frac{\psi_4+\Delta_\psi(2\omega)}{c_{44}^{E0}+\Delta_C(2\omega)}N^2(\omega)N(2\omega)K^{(2)},
\end{eqnarray}
 where
\begin{eqnarray*}
&&K^{(2)}=\left[(1-\beta\frac{J-K}{4}\lambda_1)^2+\beta^2(\frac{J-K}{4})^2\lambda_2^2\right](\xi^{(0)}\varphi_3-\sigma^{(0)}\lambda_2)\\
&& \qquad
+2\beta\frac{J-K}{4}\lambda_2[1-\beta\frac{J-K}{4}\lambda_1](\xi^{(0)}\lambda_2-\sigma^{(0)}\varphi_3).
\end{eqnarray*}
As one can easily verify, $\chi_{111}^{(2)}(\omega)$ and
$d_{114}^{(2)}(\omega)$ are different from zero only at
$\xi^{(0)}\neq 0$ (non-zero polarization), i.e. in the
ferroelectric phase or in presence of an external bias field.

In the similar way we find the third-order susceptibility. Cubic
in $E_{1t}$ kinetic equations for the spin variables read
\begin{eqnarray}
  &&-\xi^{(3)} + c_2^+(\gamma^{(3)}+\beta\lambda c S^{(3)}+\delta^{(3)})+
  c_2^-(\gamma^{(3)}+\beta\lambda c S^{(3)}-\delta^{(3)})\nonumber\\&&{}+
  2c_3^+(\gamma^{(1)}+\beta\lambda c S^{(1)}+\delta^{(1)})(\gamma^{(2)}+\beta\lambda c S^{(2)}+\delta^{(2)})+
  2c_3^-(\gamma^{(1)}+\beta\lambda c S^{(1)}-\delta^{(1)})(\gamma^{(2)}+\beta\lambda c S^{(2)}-\delta^{(2)})\nonumber\\
 &&{} + c_4^+(\gamma^{(1)}+\beta\lambda c S^{(1)}+\delta^{(1)})^3+c_4^-(\gamma^{(1)}+\beta\lambda c S^{(1)}-\delta^{(1)})^3  =0,\nonumber \\
&&-\sigma^{(3)}+ c_2^+(\gamma^{(3)}+\beta\lambda c
S^{(3)}+\delta^{(3)})-
  c_2^-(\gamma^{(3)}+\beta\lambda c S^{(3)}-\delta^{(3)})\nonumber\\&& {} +
  2c_3^+(\gamma^{(1)}+\beta\lambda c S^{(1)}+\delta^{(1)})(\gamma^{(2)}+\beta\lambda c S^{(2)}+\delta^{(2)})-
  2c_3^-(\gamma^{(1)}+\beta\lambda c S^{(1)}-\delta^{(1)})(\gamma^{(2)}+\beta\lambda c S^{(2)}-\delta^{(2)})\nonumber\\
 &&{} + c_4^+(\gamma^{(1)}+\beta\lambda c S^{(1)}+\delta^{(1)})^3-c_4^-(\gamma^{(1)}+\beta\lambda c S^{(1)}-\delta^{(1)})^3  =0,
\end{eqnarray}
with
\[
c_4^\pm=\frac1{48}\left[-2+\cosh \frac{\gamma^{(0)}+\beta\lambda c
S^{(0)}\pm\delta^{(0)}}{2}\right]
\frac{1}{\cosh^4\frac{\gamma^{(0)}+\beta\lambda c
S^{(0)}\pm\delta^{(0)}}{2}}.
\]
Following the same procedure, we obtain the third order dynamic
susceptibility
\begin{eqnarray}
\label{third}
&&\chi_{1111}^{(3)}(\omega)=\frac1{6\eps_0}\dthree{P_1^{(3)}}{E_{1t}}=-\frac
{[\mu_1'(\omega)N(\omega)]^3\mu_1'(3\omega)N(3\omega)}{v\eps_0}\left[N(2\omega)K^{(3)}_1(2\omega)+K^{(3)}_2\right].
\end{eqnarray}
Notations introduced here are given in Appendix.

\section{Numerical analysis}
\label{NA}

The found above dynamic characteristics of Rochelle salt  are
expressed via the equilibrium values of the order parameters
$\xi^{(0)}$, $\sigma^{(0)}$, $S^{(0)}$ and the strains
$\eps_i^{(0)}$ ($i=1-4$). Those quantities are calculated by
finding extrema of the thermodynamic potential (\ref{pot}) and
using Eqs. (\ref{2.4a}). The values of the parameters of the
modified Mitsui model $J_0$, $K_0$, $\Delta_0$, $\psi_{3i}^\pm$,
$\psi_4$, $c_{ij}^{E0}$ and others were chosen in
\cite{our-diagonal,our-comp} by fitting the theoretical pressure
dependences of the transition temperatures, as well as the
temperature dependences of several dielectric, piezoelectric, and
elastic characteristics to experimental data. In particular, the
major criterion of the fitting was to get $T_{\rm C2}=297$~K and
$T_{\rm C1}=255$~K at ambient pressure. The values of all these
parameters except for $\psi_4$ and $c_{44}^{E0}$ remain unchanged
and can be found in \cite{our-diagonal,our-comp}.

\begin{figure}[htb]
\centerline{\includegraphics[width=0.4\textwidth]{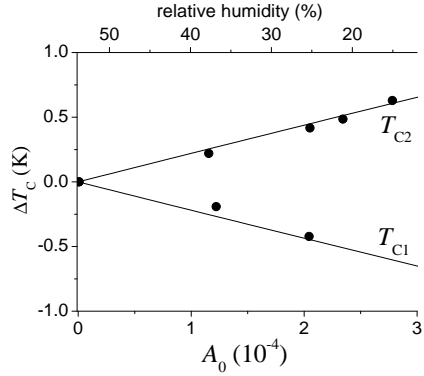}}
\caption{Dependence of the transition temperatures on the
concentration of switchable defects ($c\sim A_0$) and on the
storage atmosphere humidity. Lines: the present theory; $E_b=0$.
Symbols: experimental points of \cite{UnruhSailer}. } \label{Tc-A}
\end{figure}

Inclusion of the interactions with the defect dipoles into the
model alters the transition temperatures in the system, increasing
$T_{\rm C2}$, decreasing $T_{\rm C1}$, and widening the
ferroelectric phase, which is in agreement with experiment
\cite{UnruhSailer} (see fig.~\ref{Tc-A} and the discussion
thereof). Both the constant bias field $E_b$ of the rigid dipoles
and the switchable field of the relaxing dipoles act in this way.
Since the experimentally observed values of $T_{\rm C2}$ and
$T_{\rm C1}$ correspond to real crystals, in which defects are
unavoidable, we have to tweak slightly some parameters of the
model in such a way that the theory would yield $T_{\rm C2}<297$~K
and $T_{\rm C1}>255$~K  for a perfect crystal and $T_{\rm
C2}\approx 297$~K and $T_{\rm C1}\approx 255$~K for crystals with
defects. We take $\psi_4=-748.5$~K and $c_{44}^{E0}=1.182\cdot
10^{10}$~N/m$^2$ (c.f. $\psi_4=-750$~K, and
$c_{44}^{E0}=1.180\cdot 10^{10}$~N/m$^2$ for the model without
defects \cite{our-diagonal}).

Also we need to determine the following parameters of the
defect-mediated relaxation: $E_b$, $\lambda$, $\Psi_4$, $m_1$,
$c$, $\tau_0$, and $W$. As it has been shown in the previous
section, the piezoelectric resonances do not overlap with the
dispersion region of the defect-assisted relaxation. Therefore,
the sample dimensions are irrelevant.

If $m_1$ and $\mu_1$ are the dipole moment of the defect dipoles
and host molecules, respectively, then for the dipole-dipole
interaction constants we have $J_0+K_0\sim \mu_1^2$, $\lambda\sim
m_1\mu_1$, and for the constants of the piezoelectric coupling of
the dipoles to the shear strain $\eps_4$ we have $\psi_4\sim
\mu_1$, $\Psi_4\sim m_1$. Then we can write that
\[
\frac{\lambda}{J_0+K_0}\approx\frac{\Psi_4}{\psi_4}\approx\frac{m_1}{\mu_1}.
\]
After such a substitution, the final expressions for the
susceptibilities and piezoelectric coefficients contain only a
single combination $c\lambda^2$, instead of the four parameters
$\lambda$, $\Psi_4$, $m_1$, $c$. It is, however, more convenient
to use the parameter $A$ (\ref{a}), instead of $c\lambda^2$. We
need to set its value at the upper transition point (to be denoted
as $A_0$). It should also be mentioned that the values of the
susceptibilities are not very sensitive to the exact values of
$\Psi_4$ and $m_1$.

 The coefficient $A_0$ and the constant bias field $E_b$ are
determined by concentrations of switchable and rigid defects,
respectively and are, therefore, strongly dependent on the sample
prehistory, its quality, etc. So are the relaxation time constant
$\tau_0$ and the activation energy $W$. These parameters are to be
specified for each sample.

The activation energy $W$ has been experimentally found
\cite{UnruhSailer} to vary between 0.4 and 0.8~eV, depending on
the sample. To ascertain its value in each particular case, we
would need data on the susceptibility dispersion at two different
temperatures for each sample, preferably near the lower and upper
transition points, unfortunately not always available. We take
$W=0.5$~eV in all cases. At this value of $W$ the relaxation time
$\tau$ increases by two orders of magnitude on cooling from the
upper to the lower Curie temperature, in agreement with experiment
\cite{Unruh}.

The parameters $A_0$, $E_b$, and $\tau_0$ are found by fitting to
the frequency dependence of the linear susceptibility, or to the
Cole-Cole curves of susceptibility and linear piezoelectric
coefficient, or to the temperature curves of the linear and
non-linear susceptibilities. Note that at frequencies below and
above the defect-mediated dispersion, the susceptibilities do not
depend on $\tau_0$ or $W$.

Figure~\ref{Tc-A} compares the calculated dependences of the
transition temperatures on the parameter $A_0\sim c$  in absence
of rigid defects ($E_b=0$) with the experimental dependences of
$T_{\rm C1,2}$ on the humidity of the storage atmosphere. Overall,
a good quantitative agreement is obtained, although the
experimental $T_{\rm C1,2}$ vs humidity dependences are
non-linear. This discrepancy stems from the assumed here linear
dependences of $A_0$ and the switchable defect concentration $c$
on the changes in the humidity, while the experimental results
\cite{UnruhSailer} indicate some non-linearity.

\begin{figure}[htb]
\centerline{\includegraphics[width=0.45\textwidth]{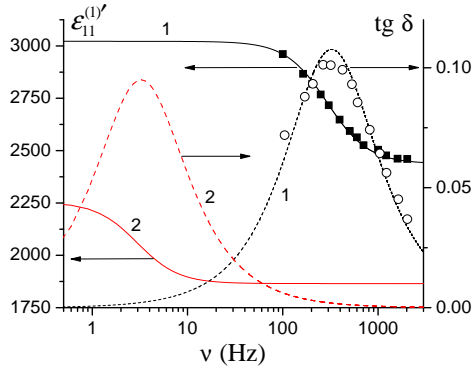}}
\caption{Frequency dependences of the real part of permittivity
and loss angle of Rochelle salt at 297.25~K (1) and 254.3~K (2).
Lines: the present theory; $A_0=8\cdot 10^{-5}$, $E_b=6.2 \cdot
10^3$~V/m, $\tau=4\cdot 10^{-4}$~s at 297.25~K and $4.4\cdot
10^{-2}$~s at 254.3~K ($\tau_0=5.95\cdot 10^{-16}$~s, $W=0.5$~eV).
Symbols: experimental points of \cite{UnruhMuser}. }
\label{MuserUnruh}
\end{figure}

\begin{figure}[htb]
\centerline{\includegraphics[width=0.75\textwidth]{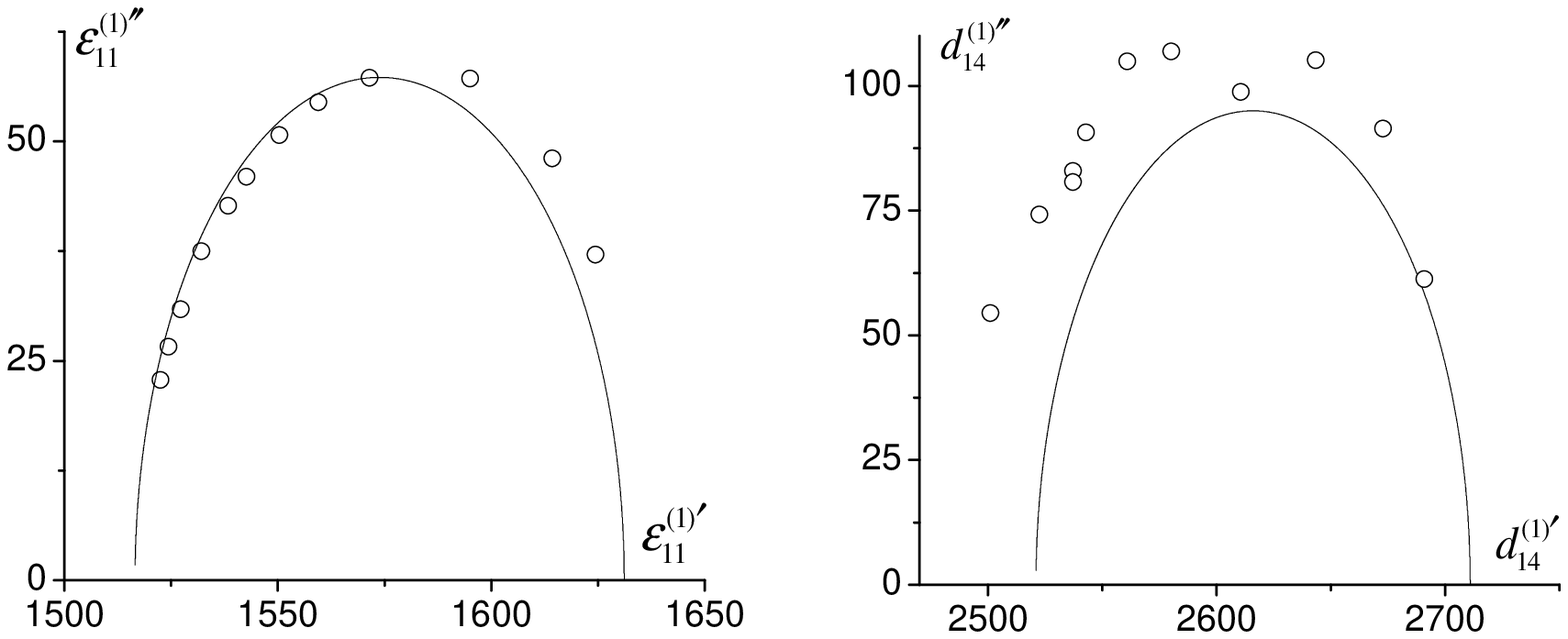}}
\caption{Cole-Cole diagrams of the linear permittivity
$\chi_{11}^{(1)}$ at 298.14~K (left) and piezoelectric coefficient
$d_{14}^{(1)}$ at 298.15~K (right). Lines: the present theory;
$A_0=4.7\cdot 10^{-5}$, $E_b=2.8 \cdot 10^3$~V/m, $\tau=6.75\cdot
10^{-4}$~s ($\tau_0=9.95\cdot 10^{-16}$~s, $W=0.5$~eV). Symbols:
experimental points of \cite{MuserSchmitt} for 298.15~K. }
\label{MuserSchmitt}
\end{figure}

The frequency variation of the linear permittivity and loss angle
of Rochelle salt just above the upper Curie temperature is shown
in fig.~\ref{MuserUnruh}. The Cole-Cole diagrams of the linear
permittivity and piezoelectric coefficients are given in
fig.~\ref{MuserSchmitt}. As one can see,  behavior of the
dielectric and piezoelectric characteristics, driven by the
dynamics of switchable defects, has a typical  relaxational
character and is well described by the present theory. The
dispersion width and the imaginary part of $d_{14}^{(1)}(\omega)$
are, however, slightly smaller than experimentally observed.
Because of the Arrhenius behavior of the relaxation time $\tau$,
the dispersion region is shifted to lower frequencies, as
temperature decreases.

\begin{figure}[htb]
\centerline{\includegraphics[width=0.45\textwidth]{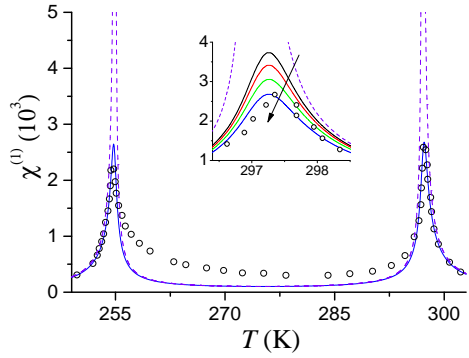} }
\caption{Temperature dependence of the real part of the linear
 permittivity $\chi_{11}^{(1)}$ of Rochelle salt at 1~kHz.
Insert: $\chi_{11}^{(1)}$ vs $T$  in the vicinity  of the upper
transition point at 0~Hz, 0~Hz, 100~Hz, 200~Hz, and 1~kHz
(frequency increases along the  arrow). Lines: the present theory;
solid lines: $A_0=1.1\cdot 10^{-4}$, $E_b=4.5 \cdot 10^3$~V/m,
$\tau_0=5.95\cdot 10^{-16}$~s, $W=0.5$~eV; dashed lines: $A=0$,
$E_b=0$ (an ideal crystal). Symbols: experimental points of
\cite{Miga} for 1~kHz. } \label{chi1}
\end{figure}

\begin{figure}[htb]
\centerline{\includegraphics[width=0.45\textwidth]{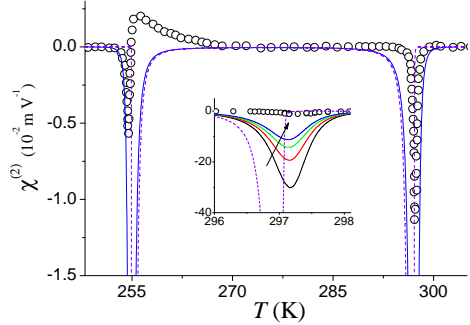} }
\caption{Temperature dependence of the real part of the second
order  permittivity $\chi_{11}^{(2)}$ of Rochelle salt at 1~kHz.
Insert: $\chi_{111}^{(2)}$ vs $T$  in the vicinity  of the upper
transition point at 0~Hz, 0~Hz, 100~Hz, 200~Hz, and 1~kHz
(frequency increases along the  arrow). Lines and the values of
$A$, $E_b$, $\tau_0$, and $W$ are the same as in fig.~\ref{chi1}.
Symbols: experimental points of \cite{Miga} for 1~kHz. }
\label{chi2}
\end{figure}

\begin{figure}[htb]
\centerline{\includegraphics[width=0.45\textwidth]{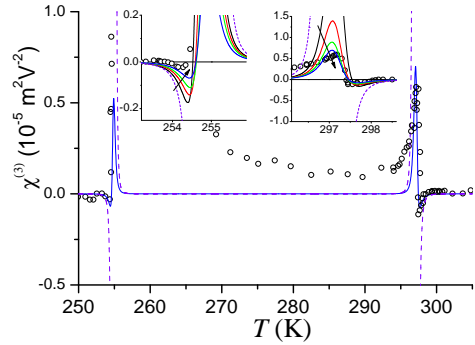}}
\caption{Temperature dependence of the real part of the  third
order permittivity $\chi_{11}^{(3)}$ of Rochelle salt at 1~kHz.
Inserts: $\chi_{1111}^{(3)}$ vs $T$  in the vicinities  of the
lower transition point at 0~Hz, 0~Hz, 0.8~Hz, 1.5~Hz, 10~Hz and of
the upper transition point at 0~Hz, 0~Hz, 100~Hz, 200~Hz, and
1~kHz (frequency increases along the arrows). Lines and the values
of $A$, $E_b$, $\tau_0$, and $W$ are the same as in
fig.~\ref{chi1}. Symbols: experimental points of \cite{Miga} for
1~kHz.  } \label{chi3}
\end{figure}

In figures~\ref{chi1}-\ref{chi3} we plot the temperature
dependences of the linear, second, and third order dynamic
permittivities of Rochelle salt at different frequencies. The
theory is compared to the experimental data of \cite{Miga}, which
have been obtained simultaneously for all three susceptibilities
and have, therefore, be described consistently, using a single set
of $A_0$, $E_b$, $\tau_0$, $W$.  We do not expect to obtain any
quantitative description of experiment in the ferroelectric phase,
where the domain contributions, not included into our model, are
predominant.

The dashed lines correspond to static susceptibilities of a
perfect crystal without defects ($A_0=0$, $E_b=0$). Their behavior
is typical for ferroelectrics with the second-order phase
transitions and agrees with the predictions of the Landau theory
\cite{Dec}, where all three susceptibilities actually diverge at
the Curie temperature. A quantitative agreement with experimental
data in the transition regions, however, is poor.

On the other hand, the dynamic susceptibilities, calculated for a
crystal with defects ($A_0\neq0$, $E_b\neq0$, solid lines), are in
a much better agreement with experiment. For $\chi_{11}^{(1)}$ and
$\chi_{1111}^{(3)}$ a very good fit is obtained, especially near
$T_{\rm C2}$, whereas for $\chi_{111}^{(2)}$ the agreement is
still not satisfactory. The non-zero values of the second order
susceptibility $\chi_{111}^{(2)}$ in the paraelectric phases are
caused by the bias field of the rigid defects $E_b$. The observed
smearing of the anomalies is caused both by the bias field  $E_b$
and by the relaxational dispersion owing to the switchable
defects. At temperatures far from the transition points the
dispersion width is small (see Eq. (\ref{dispersionwidth})), and
the influence of the constant bias field is minor; hence, the
susceptibilities of crystals with and without defects are
practically the same.

The best agreement with  experimental data for different samples,
as illustrated in figs.~\ref{lowres}-\ref{chi3}, is obtained when
the bias field $E_b$ is in the range $10^{3}\div 10^4$~V/m, which
seem to be reasonable values, and when $A_0$ is in the range
$4\cdot 10^{-5}\div3\cdot 10^{-4}$, which accords well with the
results of \cite{Unruh,UnruhSailer}.

\section{Concluding remarks}
We propose a model that considers interactions of the ordering
dipoles of a ferroelectric with dipoles, associated with crystal
defects that can be switched by the external electric field. As an
example of the ferroelectric, the Rochelle salt is taken, for
which the deformable pseudospin Mitsui model is used. The
calculated shifts of the transition temperatures with increasing
defect concentration are in a good agreement with experimental
observations. Assuming the Glauber-type kinetics of both ordering
and defect pseudospins, we calculate the linear, second, and third
order dynamic susceptibilities and piezoelectric coefficients of
the system.

The presented general scheme of taking into account the
defect-mediated relaxation can be easily generalized to other
order-disorder ferroelectrics, described by pseudospin models
(e.g. of the KH$_2$PO$_4$ family).

Dispersion of the dynamic characteristics below 1~kHz, caused by
dynamics of the relaxing defects, is described; a satisfactory
agreement with experiment is obtained. The influence of the
defect-mediated dynamics on the physical characteristics of
Rochelle salt is essential in the vicinities of the transition
points, whereas far from these temperatures the role of this
dynamics is minor. Behavior of the linear and non-linear
susceptibilities close to $T_{\rm C1,2}$ cannot be satisfactorily
described without taking into account of this dynamics and of the
constant bias field of the rigid defects.

Note that the calculations were performed within the mean field
approximation; in particular, spatial fluctuations of defect
concentration were neglected. For instance, it might be expected
that the concentration of water vacancies/interstitials is larger
in the near-surface regions of crystal samples. Possibly this is
one of reasons for the remaining discrepancies between theory and
experiment for the second-order dielectric susceptibility
$\chi_{111}^{(2)}$.

\section*{Acknowledgement}
The author acknowledges support from the State Foundation for
Fundamental Studies of Ukraine, Project No F53.2/070.








\section*{Appendix}
Notations introduced in Eq.~(\ref{xi-sigma-eps}) are
\begin{eqnarray*}
&&F_{1}^\sigma(\omega)=-\frac{\lambda_2}{\varphi _2-\Lambda_e(\omega)\varphi_3},\\
&& \varphi _2 = 1 - \frac{\beta J}{2}\lambda _1 + \beta
^2\frac{J^2 - K^2}{16}(\lambda _1^2 - \lambda _2^2 ),\quad \varphi
_3 = \lambda _1 - \frac{\beta (J - K)}{4}(\lambda _1^2 - \lambda
_2^2 ),\\
&& \lambda _1 = 1 - (\xi^{(0)}) ^2 - (\sigma^{(0)})^2, \quad
\lambda _2 = 2\xi^{(0)}\sigma^{(0)}.
\end{eqnarray*}
The intermediate results for the order parameters, strains, and
their linear combinations are
\begin{eqnarray*}
&&\xi^{(1)}=\beta\mu_1'(\omega)N(\omega)\frac{\varphi_3}{2} E_{1t},\\
&&\gamma^{(1)}+\beta \lambda c S^{(1)}=\beta\mu_1'(\omega)N(\omega)\left[1-\frac{\beta(J-K)}{4}\lambda_1\right]E_{1t},\\
&&\sigma^{(1)}=-\beta\mu_1'(\omega)N(\omega)\frac{\lambda_2}{4}E_{1t},\\
&&\delta^{(1)}=-\beta\mu_1'(\omega)N(\omega)\frac{\beta(J-K)}{4}\lambda_2E_{1t},\\
&&\xi^{(2)}=-\frac{\beta^2}4[\mu_1'(\omega)]^2N^2(\omega)N(2\omega)K^{(2)}E_{1t}^2,\\
&&\eps^{(2)}_4=\frac{\beta^2}2[\mu_1'(\omega)]^2N^2(\omega)N(2\omega)\frac{\psi_4+\Delta_\psi(2\omega)}{v(c_{44}^{E0}+\Delta C(2\omega))}K^{(2)}E_{1t}^2,\\
&&\gamma^{(2)}+\beta \lambda c S^{(2)}=-\frac{\beta^2}2[\mu_1'(\omega)]^2N^2(\omega)N(2\omega)Z(2\omega)K^{(2)}E_{1t}^2,\\
&&\sigma^{(2)}=\frac{\beta^2}4[\mu_1'(\omega)]^2N^2(\omega)N(2\omega)K^{(2)}_\sigma E_{1t}^2,\\
&&\delta^{(2)}=\frac{\beta^2}2\frac{J-K}{4}[\mu_1'(\omega)]^2N^2(\omega)N(2\omega)K^{(2)}_\sigma
E_{1t}^2.
\end{eqnarray*}

Notations used in the expression for the third order
susceptibility (\ref{third}) are
\begin{eqnarray*}
&&K^{(3)}_1(2\omega)=-\frac18(\xi^{(0)}+\sigma^{(0)})(\lambda_1-\lambda_2)[1-\beta\frac{J-K}{4}(\lambda_1+\lambda_2)]^2[Z(2\omega)K^{(2)}-\beta\frac{J-K}{4}K^{(2)}_\sigma]\\
&& \qquad
-\frac18(\xi^{(0)}-\sigma^{(0)})(\lambda_1+\lambda_2)[1-\beta\frac{J-K}{4}(\lambda_1-\lambda_2)]^2[Z(2\omega)K^{(2)}+\beta\frac{J-K}{4}K^{(2)}_\sigma];
\\
&&
K^{(3)}_2=-\frac{1}{48}[2-3(\lambda_1+\lambda_2)](\lambda_1+\lambda_2)\left[1-\beta\frac{J-K}{4}(\lambda_1+\lambda_2)\right]^4\\
&&\qquad
-\frac{1}{48}[2-3(\lambda_1-\lambda_2)](\lambda_1-\lambda_2)\left[1-\beta\frac{J-K}{4}(\lambda_1-\lambda_2)\right]^4,
\end{eqnarray*}
where \begin{eqnarray*}
&&K^{(2)}_\sigma=\left[(1-\beta\frac{J-K}{4}\lambda_1)^2+\beta^2(\frac{J-K}{4})^2\lambda_2^2\right][\xi^{(0)}\lambda_2-\sigma^{(0)}\lambda_1+
\sigma^{(0)} Z(2\omega) (\lambda_1^2-\lambda_2^2)]\\
&& \qquad
-2\beta\frac{J-K}{4}\lambda_2[1-\beta\frac{J-K}{4}\lambda_1][\sigma^{(0)}\lambda_2-\xi^{(0)}\lambda_1+\xi^{(0)}
Z(2\omega)(\lambda_1^2-\lambda_2^2)];
\\
&&
Z(2\omega)=\beta\frac{J+K}{4}+\Lambda_e(2\omega)+\Lambda(2\omega).
\end{eqnarray*}

\end{document}